\documentclass[aps,prd,superscriptaddress,amsmath,amssymb,amsthm,eqsecnum,nofootinbib,twocolumn, preprintnumbers,nolongbibliography,10pt]{revtex4-2}
\usepackage{url}
\usepackage[colorlinks=true,
    linkcolor=blue,
    citecolor=blue,
    urlcolor=blue]{hyperref}
 \usepackage{amsmath}
\usepackage{graphicx}
\usepackage{epstopdf}
\usepackage{float}
\usepackage{subcaption}
\usepackage{caption}

\usepackage{color}
\usepackage[T1]{fontenc}
\usepackage[utf8]{inputenc}
\usepackage[toc,page]{appendix}
\usepackage[usenames,dvipsnames]{xcolor}
\usepackage[normalem]{ulem}

\makeatletter
\providecommand*{\diff}%
{\@ifnextchar^{\DIfF}{\DIfF^{}}}
\def\DIfF^#1{%
\mathop{\mathrm{\mathstrut d}}%
\nolimits^{#1}\gobblespace}
\def\gobblespace{%
\futurelet\diffarg\opspace}
\def\opspace{%
\let\DiffSpace\!%
\ifx\diffarg(%
\let\DiffSpace\relax
\else
\ifx\diffarg[%
\let\DiffSpace\relax
\else
\ifx\diffarg\{%
\let\DiffSpace\relax
\fi\fi\fi\DiffSpace}

\usepackage{xcolor}
\usepackage{scalerel}

\usepackage{tikz}
\usetikzlibrary{svg.path}
\definecolor{darkmagenta}{HTML}{8B008B}
\definecolor{orcidlogocol}{HTML}{A6CE39}
\tikzset{
  orcidlogo/.pic={
    \fill[orcidlogocol] svg{M256,128c0,70.7-57.3,128-128,128C57.3,256,0,198.7,0,128C0,57.3,57.3,0,128,0C198.7,0,256,57.3,256,128z};
    \fill[white] svg{M86.3,186.2H70.9V79.1h15.4v48.4V186.2z}
                 svg{M108.9,79.1h41.6c39.6,0,57,28.3,57,53.6c0,27.5-21.5,53.6-56.8,53.6h-41.8V79.1z M124.3,172.4h24.5c34.9,0,42.9-26.5,42.9-39.7c0-21.5-13.7-39.7-43.7-39.7h-23.7V172.4z}
                 svg{M88.7,56.8c0,5.5-4.5,10.1-10.1,10.1c-5.6,0-10.1-4.6-10.1-10.1c0-5.6,4.5-10.1,10.1-10.1C84.2,46.7,88.7,51.3,88.7,56.8z};
  }
}

\newcommand\orcidicon[1]{\href{https://orcid.org/#1}{\mbox{\scalerel*{
\begin{tikzpicture}[yscale=-1, scale=0.05, transform shape]
\pic{orcidlogo};
\end{tikzpicture}
}{|}}}}

\begin{document}


\title{From Accelerating Rotating Black Holes to a Swirling Universe}

\author{Jos\'e Barrientos\,$^{\orcidicon{0000-0003-3445-8151}}$}
\email{barrientos@math.cas.cz}
\affiliation{Institute of Mathematics of the Czech Academy of Sciences, {\v Z}itn{\'a} 25, 115 67 Praha 1, Czech Republic}
\affiliation{Vicerrector\'ia de Investigaci\'on y Postgrado, Universidad de La Serena, La Serena 1700000, Chile}

\author{Adolfo Cisterna\,$^{\orcidicon{0000-0002-6949-7882}}$}
\email{adolfo.cisterna@mff.cuni.cz}
\affiliation{Sede Esmeralda, Universidad de Tarapac{\'a}, Avenida Luis Emilio Recabarren 2477, Iquique, Chile}
\affiliation{Institute of Theoretical Physics, Faculty of Mathematics and Physics, Charles University, V Hole{\v s}ovi{\v c}k{\' a}ch 2, 180 00 Praha 8, Czech Republic}

\begin{abstract}
We demonstrate that the swirling universe can be recovered through a coordinated horizon-degeneration and scaling procedure applied to the Kerr C-metric. The construction exploits the horizon structure of accelerating, rotating black holes, which allows their inner and outer horizons to coalesce while simultaneously approaching the acceleration horizon. By magnifying the near-acceleration-horizon region faster than the black hole horizons approach it, the pair of accelerating, rotating black holes is pushed to an infinite proper distance, while a finite gravitomagnetic twist survives. The resulting geometry reproduces precisely the swirling universe, providing a physical interpretation of its intrinsic rotation as the gravitational remnant of rotating black holes sent to infinity.
\end{abstract}
\maketitle

\section{Introduction}

The construction of exact black hole solutions to Einstein's equations typically assumes a constant-curvature background, such as Minkowski, de Sitter, or anti-de Sitter spacetime. However, more general gravitational environments are both physically motivated and worth exploring. Astrophysical black holes are often surrounded by accreting matter, external electromagnetic fields, and other sources of distortion that can significantly alter their surroundings \cite{Bicak:2000zk}.

A prototypical example of a nonconstant-curvature background is the Melvin--Bonnor spacetime \cite{Melvin:1963qx, Bonnor_1954}, which describes a cylindrically symmetric magnetic field coupled to its own gravitational field. The magnetic field is approximately uniform near the symmetry axis and decays with increasing cylindrical distance. 

This algebraically special (Petrov type D) spacetime has served as a natural setting for investigating black holes immersed in external electromagnetic fields \cite{Melvin:1965zza, Melvin1966, Ernst:1976bsr, ErnstCmetric, Gibbons:2013yq, BenAchour:2026zui, Barrientos:2026kdl, Barrientos:2026shy}. Unlike asymptotically flat geometries, the Melvin--Bonnor background exhibits a nontrivial asymptotic structure, approaching a Levi-Civita geometry at large cylindrical distances \cite{Stephani:2003tm}.

Nontrivial gravitational backgrounds, however, need not be supported by matter fields. A particularly interesting vacuum example is the swirling spacetime, originally introduced by Harrison \cite{Harrison} as a twisted, purely gravitational counterpart of the Melvin--Bonnor universe. The solution was subsequently revisited by Garfinkle and Melvin \cite{Garfinkle:1994wk}, who constructed a configuration combining both backgrounds and was later rediscovered by Contopoulos et al. \cite{Contopoulos:2015wra}, who obtained a Schwarzschild black hole immersed in the swirling geometry. Since then, its geometric and physical properties have been extensively investigated \cite{Astorino:2022aam, Barrientos:2024pkt,DiPinto:2025yaa, Cisterna:2023uqf, Barrientos:2026euk, Capobianco:2023kse, Capobianco:2024jhe, Moreira:2024sjq, Barrientos:2024umq}.

The swirling spacetime is a vacuum geometry with an intrinsic gravitomagnetic structure whose rotation is independent of the angular momentum of any compact source embedded within it. In cylindrical coordinates $\{t,\rho,z,\phi\}$, this geometry, commonly referred to as the swirling universe, 
takes the form
\begin{equation}
\begin{aligned}
\mathrm{ds}^2&=\left(1+j^2\rho^4\right)\left(-\diff t^2+\diff\rho^2+\diff z^2\right)\\
&\quad +\frac{\rho^2}{1+j^2\rho^4}\left(\diff\phi+4jz\diff t\right)^2,
\label{swirling_intro}
\end{aligned}
\end{equation}
where $j$ controls the strength of the background rotation. This regular vacuum solution is of Petrov type D, belongs to the Kundt class, and is free of curvature singularities, Misner strings, and regions containing closed timelike curves. Its characteristic feature is a non-uniform frame dragging, whose angular velocity depends linearly on the axial coordinate $z$ and reverses sign across the equatorial plane $z=0$. The two hemispheres therefore counter-rotate, with the magnitude of frame dragging increasing away from the equatorial plane, despite the absence of a net dipolar angular momentum. Moreover, the stationary Killing vector $\partial_t$ can become spacelike in the asymptotic region, giving rise to noncompact ergoregions bounded by stationary-limit surfaces \cite{Astorino:2022aam, Barrientos:2024pkt}. At large cylindrical distances, the solution approaches a twisted Levi-Civita geometry, in which ordinary translational symmetry along the axis is replaced by a symmetry combining axial translations with azimuthal shifts.

The intrinsic rotation of the swirling background also leaves a distinctive imprint on test-particle dynamics, giving rise to characteristic vortex-like trajectories. While motion in the cylindrical radial direction is bounded, motion along the axial direction is generally unbounded, allowing particles to escape to infinity along the symmetry axis \cite{Capobianco:2023kse, Capobianco:2024jhe}. 

The connection between the Melvin--Bonnor and swirling backgrounds becomes particularly transparent through their solution-generating constructions. Both can be obtained from Minkowski spacetime by exploiting hidden symmetries associated with the reduction of Einstein--Maxwell theory along a spacelike axial Killing vector \cite{Ernst:1967wx, Ernst:1967by}. The Melvin--Bonnor universe follows from a Harrison transformation \cite{Harrison} belonging to the $SU(2,1)$ hidden symmetry of the electrovacuum, whereas the swirling universe is generated by the corresponding spacelike realization of the Ehlers symmetry \cite{Ehlers:1957zz} in the vacuum sector. In both cases, the transformations exploit the norm of the axial Killing vector, $g_{\phi\phi}=\rho^2$, of the Minkowski seed. While the Harrison transformation uses this quantity to generate an external magnetic field, the Ehlers transformation generates a purely gravitational, gravitomagnetic twist. This common origin explains certain similarities in their radial structure and asymptotic behavior, despite the distinct geometric properties of the resulting backgrounds.\footnote{The Harrison and Ehlers transformations can also be combined along the axial Killing direction, yielding the electromagnetic swirling universe, in which an external electromagnetic field coexists and interacts nonlinearly with the gravitomagnetic twist of the swirling background \cite{Barrientos:2024pkt}.}

A complementary manifestation of this connection is provided by planar black hole geometries. Specifically, planar Reissner--Nordstr\"om and Taub--NUT solutions yield the Melvin--Bonnor universe \cite{Gibbons:2001sx} and the swirling background \cite{Astorino:2022aam}, respectively, after double Wick rotations of the Killing coordinates and suitable coordinate and parameter redefinitions. The continuation turns the planar seed's temporal Killing direction into the resulting background's axial direction, providing a geometric counterpart to the axial realization of the Harrison and Ehlers transformations. The same procedure applied to a planar Reissner--Nordstr\"om--NUT geometry yields the electromagnetic swirling universe \cite{Barrientos:2024pkt}.

This perspective motivates an alternative construction of the Melvin--Bonnor universe as the fully backreacting gravitational counterpart of a well-known process in classical electrodynamics. In the test-field approximation, the Born solution describes the electromagnetic field generated by a pair of uniformly accelerated charges. By sending these sources to infinity while appropriately increasing their charges, a finite, homogeneous electromagnetic field survives in Minkowski spacetime.

Havrdov\'a and Krtou\v{s} \cite{Havrdova:2006gi} established the gravitational counterpart of this limiting procedure using the charged C-metric \cite{Farhoosh:1979tk, Hong:2003gx}, which describes a pair of charged, accelerating black holes. By bringing the black hole horizons to the acceleration horizon and zooming into the near-acceleration-horizon region, the black holes are pushed to infinite distance in the rescaled geometry, while a finite, self-gravitating electromagnetic field remains. The resulting spacetime is precisely the Melvin--Bonnor universe, which thus admits a natural interpretation as the electromagnetic field left behind by two accelerating, charged black holes that have been sent asymptotically far.

This naturally raises the question of whether an analogous, fully backreacting limiting procedure can recover the swirling universe as the gravitational twist left behind by rotating sources sent to infinity.

The natural candidates are twisting extensions of the C-metric, where additional rotation parameters enlarge the horizon structure of the accelerating geometry.
The common gravitomagnetic origin of the Taub--NUT and swirling spacetimes, rooted in their respective constructions via timelike and spacelike realizations of the Ehlers symmetry \cite{Ehlers:1957zz}, naturally suggests accelerating NUT geometries \cite{Barrientos:2023tqb, Astorino:2024bfl} as candidates for implementing the limiting procedure of Havrdov\'a and Krtou\v{s}. 

However, we show that their horizon structure obstructs the required degeneration. As discussed in App.~\ref{appnutlimit} for the solution of Ref.~\cite{Astorino:2024bfl}, the inner and outer black hole horizons satisfy $r_+r_-=-l^2$, where $l$ is the NUT parameter. For real $l\neq 0$, the two horizons necessarily lie on opposite sides of $r=0$, preventing them from simultaneously approaching the acceleration horizon in the positive radial domain.

We demonstrate that the required horizon configuration is instead provided by the Kerr C-metric \cite{Griffiths:2005se} spacetime, where angular momentum plays the crucial role. Indeed, its inner and outer black hole horizons satisfy $r_+r_-=a^2$, with both roots lying in the positive radial domain for the relevant parameter range. This allows us to construct a limit in which the two black hole horizons coalesce while simultaneously approaching the acceleration horizon, thereby establishing a gravitational analog of the charged C-metric-Melvin limiting procedure of Ref.~\cite{Havrdova:2006gi}. We explicitly show that, upon implementing this degeneration and appropriately scaling the near-acceleration-horizon region, the pair of accelerating, rotating black holes is effectively sent to infinity, while a finite gravitomagnetic twist survives, which is precisely the swirling universe. 

The paper is organized as follows.  In \autoref{sec2a}, we introduce the rotating C-metric and discuss the horizon structure relevant for the limiting construction.  In \autoref{sec2b}, we implement the double-horizon scaling in which the inner and outer black hole horizons approach the acceleration horizon.  In \autoref{sec2c}, after suitable coordinate transformations, we identify the resulting geometry with the swirling background.  Finally, in \autoref{secfinal}, we summarize our results and discuss possible extensions. Appendix~\ref{appnutlimit} provides details on the limiting construction in the presence of the NUT parameter and discusses the obstruction to obtaining the swirling limit.

\section{Towards the limiting geometry}
\subsection{Seed configuration: rotating C-metric}
\label{sec2a}

To implement the limiting construction, we use the Petrov type D rotating C-metric introduced in Ref.~\cite{Griffiths:2005se}. The solution is characterized by the mass parameter $m$, the Kerr rotation parameter $a$, and the acceleration parameter $\alpha$. Its appropriate extension describes a pair of accelerating Kerr black holes; the coordinates used below cover a single accelerated exterior region. In Boyer--Lindquist-type coordinates $\{t,r,x=\cos\theta,\varphi\}$,
the metric reads
\begin{align}
\mathrm{ds}^2=&\frac{1}{\Omega^2}\Bigg[
-\frac{\Delta_r}{\Sigma}
\left(\diff t-a(1-x^2)\diff\varphi\right)^2\nonumber+\Sigma\left(\frac{\diff r^2}{\Delta_r}
+\frac{\diff x^2}{\Delta_x}\right)\nonumber\\
&\qquad+\frac{\Delta_x}{\Sigma}
\left(a\diff t-(r^2+a^2)\diff\varphi\right)^2\Bigg],
\label{metric_rotating_C}
\end{align}
where
\begin{align}
\Omega&=1-\alpha r x,\\
\Sigma&=r^2+a^2x^2,\\
\Delta_r&=(1-\alpha^2r^2)(r^2-2mr+a^2),\label{Dr_rotating_C}\\
\Delta_x&=(1-x^2)\left(1-2\alpha mx+\alpha^2a^2x^2\right).\label{Dx_rotating_C}
\end{align}
The radial function factorizes as
\begin{equation}
\Delta_r=(1-\alpha^2r^2)(r-r_+)(r-r_-),
\end{equation}
with
\begin{equation}
r_\pm=m\pm\sqrt{m^2-a^2}, \qquad r_+r_-=a^2.
\label{roots_rotating_C}
\end{equation}
The roots $r_\pm$ correspond to the outer and inner black hole horizons, while $r=\pm1/\alpha$ are the acceleration roots. We reserve $r_{\rm A}$ for the positive acceleration horizon and consider the subextremal branch
\[
\alpha>0,\qquad m>|a|>0,\qquad
0<r_-<r_+<r_{\rm A}.
\]
The exterior region of interest is $r_+<r<r_{\rm A}$, with
$-1<x<1$. These conditions ensure $\Delta_r>0$, $\Omega>0$, and
\[
\Delta_x=(1-x^2)(1-\alpha r_+x)(1-\alpha r_-x)>0,
\]
throughout this region. Unlike the nonrotating vacuum seed, the rotating geometry possesses two positive black hole horizon roots that can be driven toward the same acceleration horizon. This is the horizon structure required by the limiting procedure below.

\subsection{Double-horizon limit}
\label{sec2b}

The aim is to isolate the gravitational field of the accelerating black holes while sending the compact sources to infinite distance. For the construction considered here, we require a family whose inner and outer black hole horizons can coalesce while approaching the same acceleration horizon. The latter bounds the causal domain of accelerated observers rather than the black hole region and provides the reference surface around which we magnify the geometry. The parameter degeneration is accompanied by a coordinate scaling that approaches the acceleration horizon faster than the black hole horizons do. Their locations then recede to infinity in the rescaled coordinates, while a finite background field survives. We also verify below that the proper axial separation diverges, giving this interpretation a geometric basis.

We implement a double-horizon scaling analogous to that employed in the charged C-metric-Melvin limit \cite{Havrdova:2006gi}. Keeping $\alpha>0$ fixed, we introduce a small parameter $\epsilon$ and two constants $\lambda_\pm$, and write
\begin{equation}
\alpha r_\pm=1+\lambda_\pm\epsilon,
\qquad
p=\lambda_+\lambda_-,
\label{horizon_scaling}
\end{equation}
where
\[
\epsilon\to0^+,\qquad \lambda_-<\lambda_+<0.
\]
For sufficiently small $\epsilon$, this choice preserves the ordering $0<r_-<r_+<r_{\rm A}$ and implies $p>0$. The constants $\lambda_\pm$ specify the path through parameter space; the parameters $m$ and $a$ vary along this path rather than being held fixed.

Using
\begin{equation}
r_++r_-=2m, \qquad r_+r_-=a^2,
\end{equation}
we obtain the exact relations
\begin{equation}\label{m_scaling}
\begin{aligned}
\alpha m &= 1+\frac{\lambda_++\lambda_-}{2}\epsilon,\\
\alpha^2a^2&=1+(\lambda_++\lambda_-)\epsilon +p\epsilon^2.
\end{aligned}
\end{equation}
In particular,
\[
m^2-a^2=\frac{(\lambda_+-\lambda_-)^2}{4\alpha^2}\epsilon^2,
\]
so the black holes remain nonextremal for every $\epsilon>0$ in the chosen family. In the limit,
\begin{equation}
\alpha m\longrightarrow1, \qquad
\alpha^2a^2\longrightarrow1,\qquad m^2-a^2\rightarrow0,
\label{critical_rotating_C}
\end{equation}
and both black hole horizons coalesce while approaching
\begin{equation}\label{acchor}
r_{\rm A}=\frac{1}{\alpha}.
\end{equation}

The critical parameter values are not substituted before the coordinate scaling. Instead, along the same family of nonextremal geometries, we simultaneously magnify an increasingly narrow region around the acceleration horizon by introducing
\begin{equation}
\alpha r =1+p\epsilon^2v.
\label{radial_scaling}
\end{equation}
At fixed $v$, the observation point approaches $r_{\rm A}$ at order
$\epsilon^2$, whereas the black hole horizons approach it at order
$\epsilon$. 

Using Eqs.~\eqref{horizon_scaling} and \eqref{radial_scaling}, we find
\begin{equation}\label{Dr_limit}
\Delta_r =-\frac{2p^2}{\alpha^2}\epsilon^4v+\mathcal{O}(\epsilon^5).
\end{equation}
After their powers of $\epsilon$ are extracted, the two black hole factors contribute only constants at leading order. The dependence on $v$ therefore comes entirely from the simple acceleration horizon factor. This gives the linear radial function needed for the Rindler-like sector of the limiting metric. App.~\ref{appnutlimit} contrasts this construction with the single-black-hole-horizon scaling of the accelerating NUT seed considered there.

The remaining functions simplify as well. From Eqs.~\eqref{m_scaling}--\eqref{radial_scaling}, we obtain
\begin{align}
\Omega&=1-x+\mathcal{O}(\epsilon^2),\\
\Sigma&=\frac{1+x^2}{\alpha^2}+\mathcal{O}(\epsilon),\\
\Delta_x&=(1-x)^3(1+x)+\mathcal{O}(\epsilon).
\label{Dx_limit_rot}
\end{align}
These expansions describe the limit at fixed finite $v$ and fixed interior $x$. Correlated approaches to the angular endpoints require separate consideration, as illustrated by the stationary-limit surfaces below. $ \alpha$ fixes the scale of the resulting geometry, while the sign of $a$ determines the orientation of the surviving rotation.

To establish the geometric separation of the sources, consider the proper length along the $x=-1$ axis on a constant-$t$ slice between the outer black hole horizon and the acceleration horizon:
\begin{equation}
\mathcal L(\epsilon)=\int_{r_+}^{r_{\rm A}}
\frac{\sqrt{r^2+a^2}}
{(1+\alpha r)\sqrt{\Delta_r}}\,\diff r.
\end{equation}
Set $b=-\lambda_+$ and $c=-\lambda_-$, so that $0<b<c$, and introduce $u=(1-\alpha r)/\epsilon$. The integration interval becomes $0\leq u\leq b$, and the radial function takes the form
\begin{equation}
\Delta_r=\frac{2\epsilon^3}{\alpha^2}
u(b-u)(c-u)\,[1+\mathcal O(\epsilon)].
\end{equation}
Consequently,
\begin{equation}
\mathcal L(\epsilon)\sim
\frac{1}{2\alpha\sqrt{\epsilon}}
\int_0^b\frac{\diff u}{\sqrt{u(b-u)(c-u)}}.
\end{equation}
The remaining integral is finite and positive for $0<b<c$. Therefore $\mathcal L\propto\epsilon^{-1/2}\to\infty$: the axial proper distance diverges, not merely the horizon coordinates.

We still need to adapt the Killing coordinates to the acceleration horizon and perform the time rescaling required for a nondegenerate metric. We carry out this step next and identify the surviving background.

\subsection{Swirling spacetime}
\label{sec2c}

We evaluate the metric on the exterior side of the acceleration horizon, where $\Delta_r>0$, by setting
\begin{equation}
v=-\nu,
\qquad
\nu>0.
\end{equation}
Before taking the limit, we must adapt the angular coordinate to the rotation of the acceleration horizon. The behavior $\Delta_r=\mathcal O(\epsilon^4)$ requires a time rescaling of order $\epsilon^{-2}$. Keeping the original azimuthal coordinate fixed under this rescaling would make $a\diff t-(r^2+a^2)\diff\varphi$ divergent, while its square has a finite, nonzero prefactor at fixed off-axis points. We therefore pass to a co-rotating frame, aligning the time direction with the null generator of the acceleration horizon,
\begin{equation}
\chi_{\rm A}=\partial_t+\Omega_{\rm A}\partial_\varphi.
\end{equation}

The angular velocity follows directly from the metric. At $r=r_{\rm A}$, the radial function vanishes and the remaining contribution in the Killing sector is proportional to the square of
\begin{equation}
a\diff t-(r_{\rm A}^2+a^2)\diff\varphi.
\end{equation}
The generator is therefore null when
\begin{equation}
a-(r_{\rm A}^2+a^2)\Omega_{\rm A}=0,
\end{equation}
which gives
\begin{equation}
\Omega_{\rm A}=\frac{a}{r_{\rm A}^2+a^2}
=\frac{\alpha^2a}{1+\alpha^2a^2}.
\end{equation}
We use this angular velocity at each finite $\epsilon$, introducing
\begin{equation}
\varphi=\Phi+\Omega_{\rm A}t, \label{corotating_rotC}
\end{equation}
so that
\begin{equation}
\left.\partial_t\right|_{\Phi}
=\partial_t+\Omega_{\rm A}\partial_\varphi
=\chi_{\rm A}.
\end{equation}
The time direction to be rescaled is thus adapted to the horizon generator rather than to the original stationary Killing vector. The cancellation needed for the limit is explicit in
\begin{equation}
\begin{aligned}
a\diff t-(r^2+a^2)\diff\varphi
=-\Omega_{\rm A}(r^2-r_{\rm A}^2)\diff t&-(r^2+a^2)\diff\Phi.
\end{aligned}
\end{equation}
The coefficient of $\diff t$ vanishes at the acceleration horizon. Using Eq.~\eqref{radial_scaling} together with $v=-\nu$, we find
\begin{equation}
r^2-r_{\rm A}^2
=-\frac{2p}{\alpha^2}\epsilon^2\nu
+\mathcal O(\epsilon^4).
\end{equation}
Its leading variation therefore compensates the forthcoming $\epsilon^{-2}$ time rescaling, leaving a finite rotational one-form.

We now set
\begin{equation}
t=\frac{2}{\alpha p\epsilon^2}\tau,
\label{time_scaling_rotC}
\end{equation}
and choose the branch $\alpha a\to1$, for which
\begin{equation}
\Omega_{\rm A}\longrightarrow\frac{\alpha}{2},\qquad
r^2+a^2\longrightarrow\frac{2}{\alpha^2}.
\end{equation}
Consequently,
\begin{equation}
a\diff t-(r^2+a^2)\diff\varphi
\longrightarrow-\frac{2}{\alpha^2}
\left(\diff\Phi-\nu\diff\tau\right).
\label{rotating_oneform_limit}
\end{equation}
Co-rotation removes the horizon's constant angular velocity, not the spatial variation of the rotational field. Indeed, the surviving angular connection $\mathcal A=-\nu\diff\tau$ has
\[
\diff\mathcal A=-\diff\nu\wedge\diff\tau\neq0.
\]
An angular-coordinate shift changes $\mathcal A$ only by an exact one-form and cannot eliminate this nonzero curvature. The residual coupling therefore represents a nontrivial angular fibration, which becomes the gravitomagnetic twist of the swirling background. Exact co-rotation is a convenient choice rather than a unique one: an angular velocity differing from $\Omega_{\rm A}$ by $\mathcal O(\epsilon^2)$ also yields a finite limit, but leaves an additional finite rigid-rotation term. In particular,
\[
\Omega_{\rm A}-\frac{\alpha}{2}
=-\frac{\alpha(\alpha a-1)^2}{2(1+\alpha^2a^2)}
=\mathcal O(\epsilon^2)
\]
on the selected branch. Using the exact horizon angular velocity removes this residual frame freedom.

Using Eqs.~\eqref{horizon_scaling}--\eqref{time_scaling_rotC} and taking $\epsilon\to0$, the line element becomes
\begin{align}
\mathrm{ds}^2=&\frac{1}{\alpha^2}\frac{1+x^2}{(1-x)^2}
\biggl[-2\nu\diff\tau^2+\frac{\diff\nu^2}{2\nu}+\frac{\diff x^2}{(1-x)^3(1+x)}\biggr]\nonumber\\
&+\frac{4}{\alpha^2}\frac{1-x^2}{1+x^2}
\left(\diff\Phi-\nu\diff\tau\right)^2.
\label{swirlinter}
\end{align}
The constants $\lambda_\pm$ no longer appear: they characterize how the black hole horizons approach the acceleration horizon, but not the resulting metric.

Before introducing cylindrical coordinates, we fix the azimuthal normalization. Writing
\begin{equation*}
\Delta_x=(1-x^2)P(x),\qquad
P(x)=1-2\alpha mx+\alpha^2a^2x^2,
\end{equation*}
regularity of the $x=-1$ axis requires
\begin{equation*}
\Delta\varphi=\frac{2\pi}{P(-1)},\qquad
P(-1)=1+2\alpha m+\alpha^2a^2.
\end{equation*}
We choose this normalization throughout the family. Since the co-rotating transformation \eqref{corotating_rotC} leaves the azimuthal periodicity unchanged, $\Phi$ has the same period. As $P(-1)\to4$, the limiting period is $\pi/2$; the factor $4$ in Eq.~\eqref{swirl_coordinates} then produces a $2\pi$-periodic angle in the limiting geometry.

At this stage, it is also instructive to understand the fate of the conical singularity present in the parent C-metric. Fixing the azimuthal periodicity to regularize the $x=-1$ axis leaves a conical defect at $x=+1$, whose deficit does not vanish under the horizon degeneration. Instead, the coordinate transformation leading to the swirling geometry maps the regular axis $x=-1$ to $\rho=0$, while the defective axis $x=+1$ is pushed to $\rho\to\infty$, an asymptotic end at infinite proper radial distance (see \eqref{swirl_coordinates}, where $\rho$ is introduced). The limiting geometry therefore retains only the regular axial sector, with the azimuthal normalization ensuring the standard $2\pi$ periodicity. 

We now introduce the coordinates
\begin{equation}
\begin{aligned}
\rho&=\frac{1}{\sqrt{2}\alpha}\sqrt{\frac{1+x}{1-x}},
 \quad V=\frac{\nu}{2\alpha^2},\quad
T=-\tau, & \psi=4\Phi.
\end{aligned}
\label{swirl_coordinates}
\end{equation}
The interval $-1<x<1$ maps to $0<\rho<\infty$. The regular axis $x=-1$ becomes $\rho=0$, whereas the other angular endpoint $x\to1$ is sent to transverse infinity. The metric takes the form
\begin{align}
\mathrm{ds}^2=&\left(1+j^2\rho^4\right)
\left[-2V\diff T^2+\frac{\diff V^2}{2V}+\diff\rho^2\right]
\nonumber\\
&+\frac{\rho^2}{1+j^2\rho^4}
\left(\diff\psi+4jV\diff T\right)^2,
\label{swirling_rindler}
\end{align}
where $j=2\alpha^2$. The other branch, $\alpha a\to-1$, reverses the orientation of the background and gives $j=-2\alpha^2$ with the same coordinate conventions.

Finally, introducing
\begin{equation}
\begin{aligned}
\tilde t=\sqrt{2V}\sinh T,\quad
z=\sqrt{2V}\cosh T,\quad
\tilde{\phi}=\psi-2j\tilde{t}z,
\end{aligned}
\label{rindler_to_swirl}
\end{equation}
and after dropping the tildes, Eq.~\eqref{swirling_rindler} reproduces the swirling spacetime in Eq.~\eqref{swirling_intro}. We retain the tildes below only to distinguish the canonical coordinates from the seed coordinates.

For $V>0$, the final transformation covers the wedge $z>|\tilde t|$, rather than the entire canonical coordinate domain. The swirling metric is regular across the null boundaries $z=\pm\tilde t$ and extends beyond this wedge. The construction therefore identifies a wedge of the swirling universe, whose regular continuation gives the extended background. The inherited acceleration horizon is a boost-type Killing horizon of this background, not a surviving compact-source horizon.

The scaling \eqref{radial_scaling} isolates an increasingly narrow region around the acceleration horizon, while the black hole horizons approach it according to \eqref{horizon_scaling}. Together with the divergence of the proper axial separation, this establishes the interpretation of the result as a finite gravitomagnetic field left behind by increasingly distant accelerating, rotating sources. The pair of sources is understood in the appropriate extension of the parent C-metric; the displayed calculation is performed in one of its exterior regions.

The limiting Killing fields distinguish the inherited acceleration horizon from the stationary-limit surfaces of the swirling geometry. The generator with a finite, nonzero limit is the rescaled vector
\begin{equation}
\widehat\chi_{\rm A}
=\frac{2}{\alpha p\epsilon^2}\chi_{\rm A}
=\partial_\tau,
\end{equation}
whereas $\chi_{\rm A}=(\alpha p\epsilon^2/2)\partial_\tau$ tends to zero under the same coordinate identification. Since $T=-\tau$, the rescaled generator tends to $-K$, where
\begin{equation}
\begin{aligned}
K=\partial_T
={}&z\partial_{\tilde t}+\tilde t\partial_z-2j(\tilde t^{\,2}+z^2)\partial_{\tilde\phi}.
\end{aligned}
\end{equation}
The angular component is essential: the inherited symmetry is a boost accompanied by an azimuthal shift. By contrast, the original stationary vector $k=\left.\partial_t\right|_\varphi$ satisfies
\[
k=\frac{\alpha p\epsilon^2}{2}\partial_\tau
-4\Omega_{\rm A}\partial_{\tilde\phi}
\longrightarrow-2\alpha\partial_{\tilde\phi}.
\]
It thus becomes axial, not the canonical stationary vector $\partial_{\tilde t}$. The latter belongs to the enhanced Killing algebra of the limiting background and is not obtained by a constant change of basis within the two inherited Killing fields.

In the parent geometry, the stationary-limit surface associated with $k$ obeys $\Delta_r=a^2\Delta_x$. At fixed $\nu>0$ and fixed interior $x$, the left-hand side is $\mathcal O(\epsilon^4)$, whereas the right-hand side approaches a positive, finite value. No nondegenerate off-axis stationary-limit surface associated with $k$ therefore survives at fixed coordinates. This statement does not hold uniformly at the axes. Near the surviving axis, writing $x=-1+\delta$ gives
\[
\Delta_x=[8+\mathcal O(\epsilon)]\delta
+\mathcal O(\delta^2).
\]
The parent stationary-limit condition then yields
\[
\delta=\frac{p^2\nu}{4}\epsilon^4+\mathcal O(\epsilon^5),
\qquad \rho=\mathcal O(\epsilon^2).
\]
Thus the corresponding branch collapses onto the axis, consistent with the axial limit of $k$.

The stationary-limit surfaces of the swirling background are instead defined with respect to $\partial_{\tilde t}$ and satisfy
\begin{equation}
16j^2\rho^2z^2=\left(1+j^2\rho^4\right)^2.
\end{equation}
They are not direct limits of the stationary-limit surface of the parent stationary generator. Their presence reflects the stationary structure and enhanced symmetry of the extended swirling geometry, rather than merely a constant redefinition of the original Killing basis.

\section{Final remarks}
\label{secfinal}

We have shown that the swirling spacetime can be obtained as a limiting geometry of the rotating C-metric.  The key ingredient is the presence of the inner and outer Kerr horizons, which can be driven toward the acceleration horizon simultaneously.  In the corresponding double-horizon scaling, the accelerating black hole sources are pushed to infinite distance in the rescaled geometry, while a finite rotational background survives.  After suitable coordinate transformations, we identify the resulting spacetime as the swirling universe.

This construction mirrors the well-known C-metric-Melvin limit introduced in Ref. \cite{Havrdova:2006gi}.  In the charged C-metric, the electromagnetic charge provides the pair of black hole horizons required for the limiting procedure, and the surviving field is the homogeneous Melvin background.  In the present vacuum case, angular momentum plays the analogous role: the Kerr parameter provides the inner horizon and allows the double-horizon degeneration, while the limiting geometry retains a finite swirling parameter $j$.

A natural extension is to include electromagnetic charge in the rotating C-metric.  Since the charged rotating solution has both angular momentum and electromagnetic charge, one might expect an analogous double-horizon scaling that retains both the rotational and electromagnetic sectors.  It would be interesting to determine whether the resulting geometry is the electromagnetic swirling universe, thereby extending the correspondence established here.

Another avenue is to include a cosmological constant.  Similar limiting procedures have been previously used to obtain Melvin-(A)dS backgrounds \cite{Lim:2018vbq, Barrientos:2024umq}. It remains to be explored whether accelerating Kerr-(A)dS geometries admit the double-horizon limit leading to swirling backgrounds with  cosmological constant.  More generally, applying related limiting procedures to broader classes of algebraically general accelerating spacetimes \cite{Barrientos:2023tqb, Barrientos:2023dlf} may reveal additional nontrivial background geometries and uncover connections among their distinct limiting sectors.
\acknowledgments
J.B. acknowledges the support of ANID through grant FOVI250117 and the hospitality of the Institute of Mathematics, Czech Academy of Sciences (RVO 67985840). A.C. is partially supported by FONDECYT grant 1250318. 
\appendix
\section{Obstruction in the accelerating NUT case}
\label{appnutlimit}

Given the close relation between NUT and swirling geometries, a natural first attempt is to take the swirling spacetime as a limit of the vacuum accelerating NUT solution \cite{Astorino:2024bfl}.  However, its horizon structure obstructs the required double-horizon degeneration.  The radial function can be written as
\begin{equation}
\Delta_r=(1-\alpha^2r^2)(r-r_+)(r-r_-),
\end{equation}
with $r_\pm=m\pm\sqrt{m^2+l^2}$, then $r_+r_-=-l^2$. For real $l$, the two black hole roots therefore have opposite signs and cannot simultaneously approach the same positive acceleration horizon $r_{\rm A}$. In other words, the solution cannot express extremality in $r_{\pm}$, and hence the limiting procedure, in its original form, is obstructed. 

To see this more explicitly, suppose that only the positive black hole root approaches the acceleration horizon,
\begin{equation}
\alpha r_+=1+c\epsilon^2, \qquad \alpha r=1+v\epsilon^2,
\end{equation}
with $c$ and $v$ finite as $\epsilon\rightarrow0$.  Since $r-r_+=\mathcal{O}(\epsilon^2)$ while $r-r_-=\mathcal{O}(1)$, the radial function behaves as
\begin{equation}
\Delta_r\propto \epsilon^4 v(v-c).
\end{equation}
The two zeros at $v=0$ and $v=c$ correspond, respectively, to the acceleration horizon and the black hole horizon, and both remain at a finite distance in the rescaled geometry.  Thus, the limiting two-dimensional Lorentzian sector has the generic form
\begin{equation}
\mathrm{ds}_2^2=-F(v)\diff\tau^2+\frac{\diff v^2}{F(v)}, \qquad F(v)\propto v(v-c).
\end{equation}
This is incompatible with the swirling geometry, whose corresponding sector is locally Rindler, up to a warp factor depending only on the transverse coordinate,
\begin{equation}
\mathrm{ds}_2^2=-2v\diff\tau^2+\frac{\diff v^2}{2v},
\end{equation}
and contains only the single horizon at $v=0$.  Equivalently, for a metric of the above form the two-dimensional Ricci scalar is $R_{(2)}=-F''(v)$. The quadratic function $F(v)\propto v(v-c)$ has nonvanishing curvature, whereas the linear Rindler function $F(v)=2v$ is flat.  Bringing only one black hole root toward the acceleration horizon therefore cannot reproduce the required swirling limit.

The remaining degeneration sends the two black hole roots toward opposite acceleration horizons, $r_\pm\longrightarrow\pm 1/\alpha$, which implies $m=0$ and $|\alpha l|=1$. At the critical point, the two possibilities are $\alpha l=\pm1$.  Considering the positive branch, one finds
\begin{equation}
\Delta_r= -\frac{1}{\alpha^2} (1-\alpha^2r^2)^2, \qquad \Delta_x=(1-x^2)^2.
\end{equation}
Although this degeneration is highly symmetric, the corresponding critical geometry is locally flat, $R_{\mu\nu\rho\sigma}=0.$ It therefore cannot reproduce the swirling universe, which is a Petrov type D vacuum spacetime, i.e., with a nonvanishing Weyl tensor. 
The failure of the accelerating NUT seed is thus rooted in its horizon structure rather than in a particular coordinate choice.  

\bibliography{biblio}

\end{document}